\documentclass{ws-procs9x6}            

\usepackage[compact]{titlesec} 
\usepackage{wrapfig}

\begin{document}

\begin{flushleft}
\vspace*{-1.50cm}
NJU-INP 008/19
\end{flushleft}

\title{Excited light baryons from quark-gluon-level calculations}

\author{J. Segovia$^*$}
\address{Departamento de Sistemas F\'isicos, Qu\'imicos y Naturales \\ Universidad Pablo de Olavide, E-41013 Sevilla, Spain \\ $^*$E-mail: jsegovia@upo.es}

\author{C. Chen}
\address{Institut f\"ur Theoretische Physik, Justus-Liebig-Universit\"at Gie\ss en \\ Heinrich-Buff-Ring 16, 35392 Giessen, Germany}

\author{Z.-F. Cui, Y. Lu, and C.D. Roberts}
\address{School of Physics and Institute for Nonperturbative Physics \\
Nanjing University, Nanjing, Jiangsu 210093, China}

\begin{abstract}
The task of mapping and explaining the spectrum of baryons and the structure of these states in terms of quarks and gluons is a longstanding cha\-llen\-ge in hadron physics, which is likely to persist for another decade or more. We review the progress made in this topic using a functional method based on Dyson-Schwinger equations. This framework provides a non-perturbative, Poincar\'e-covariant continuum formulation of Quantum Chromodynamics which is able to extract novel insight on baryon properties since the physics at the hadron level is directly related with the underlying quark-gluon substructure, via convolution of Green functions.
\end{abstract}


\bodymatter


\section{Introduction}

At high energies, many experiments combined with perturbative calculations have established that Quantum Chromodynamics (QCD) is a valid candidate for a theory of the strong interactions~\cite{Marciano:1977su}. However, the infrared dynamics of QCD represents a challenge of notorious complexity: the most characteristic phenomena such as dynamical generation of quark and gluon masses, formation of bound states and confinement are purely non-perturbative. 

The most plausible way to study QCD's non-perturbative regime is through detailed investigations of hadron's mass, structure and reaction properties. In fact, hadron physics is a key part of the international effort in basic science. The experimental programs performed at $B$-factories (BaBar, Belle and CLEO), at $\tau$-charm facilities (CLEO-c, BESIII), at proton--(anti-)proton colliders (CDF, D0, LHCb, ATLAS, CMS) and at fixed-target experiments (COMPASS, Hall-B at JLab) have provided a large amount of data that has been crucial for sustained progress in the field as well as the breadth and depth necessary for a vibrant research environment. Moreover, a new generation of facilities are either currently running such as JLab12, or scheduled for the next one-to-two decades as FAIR and EIC.

Many theoretical techniques have been developed in the last fifty years in order to tackle the problem of describing hadrons as bound-states of quarks and gluons. Among them, the QCD's Dyson-Schwinger Equations (QCD-DSEs) formalism provides an appealing tool for several reasons. QCD-DSEs are the quantum equations of motion derived from the QCD Lagrangian and thus represent a continuum approach which operates within a fully relativistic quantum field theory, provides access to both perturbative and non-perturbative regimes, and is able to cover the full quark-mass range between chiral limit and the heavy-quark domain. Note, however, that for QCD-connected analysis one relies upon a truncation of the infinite system of non-linear integral equations to a subset that captures the physical content and is solved explicitly, combined with the use of {\emph ans\"atze} for those Green functions that enter the equations but are not solved for. These {\emph ans\"atze} are constrained by symmetry properties, multiplicative renormalizability, perturbative limits, etc. The interested reader on QCD-DSEs applied to hadron physics can consult some existing reviews~\cite{Roberts:1994dr, Alkofer:2000wg}.


\begin{figure}[!t]
\begin{center}
\includegraphics[width=0.47\textwidth]{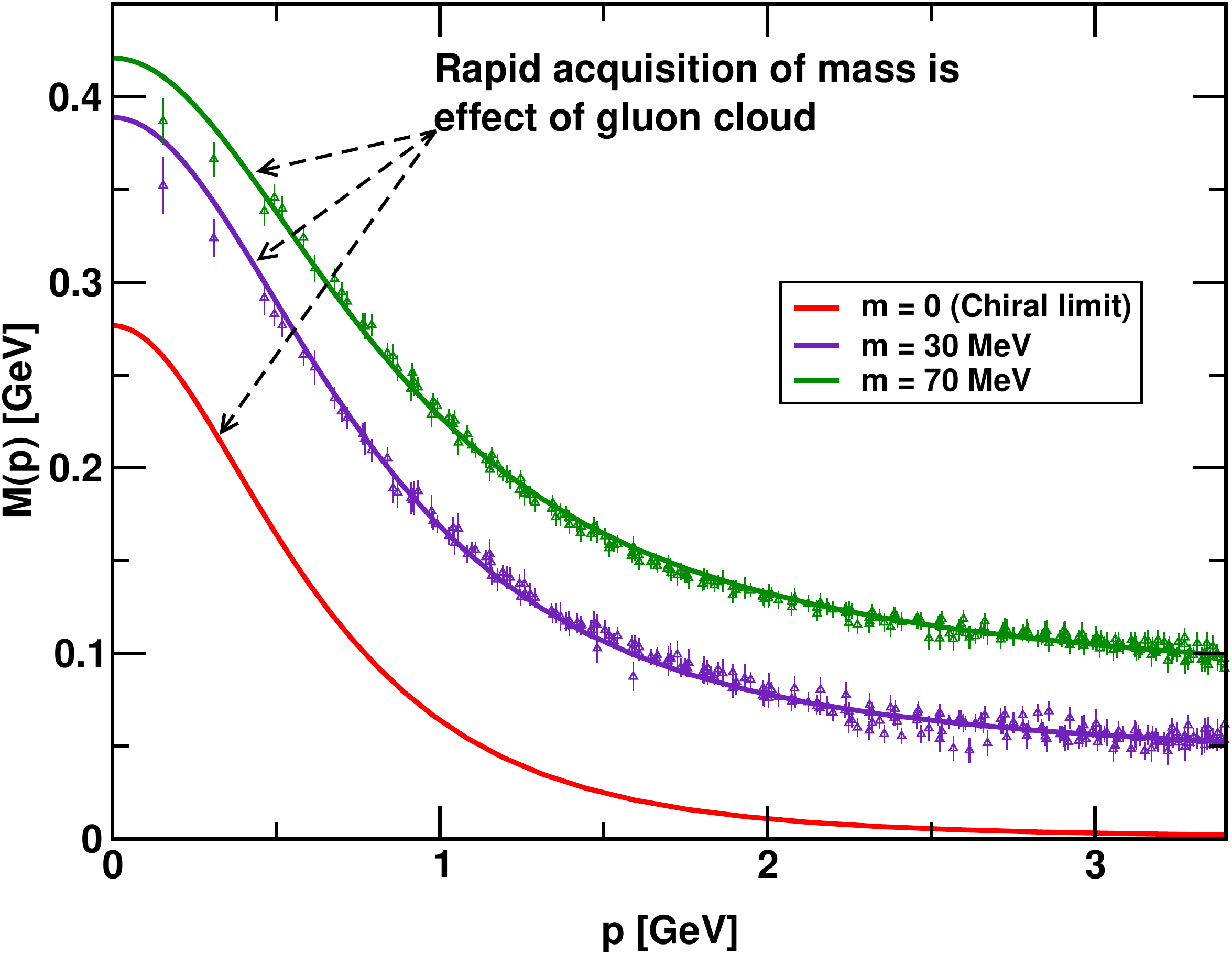}
\hspace*{0.25cm}
\includegraphics[width=0.49\textwidth]{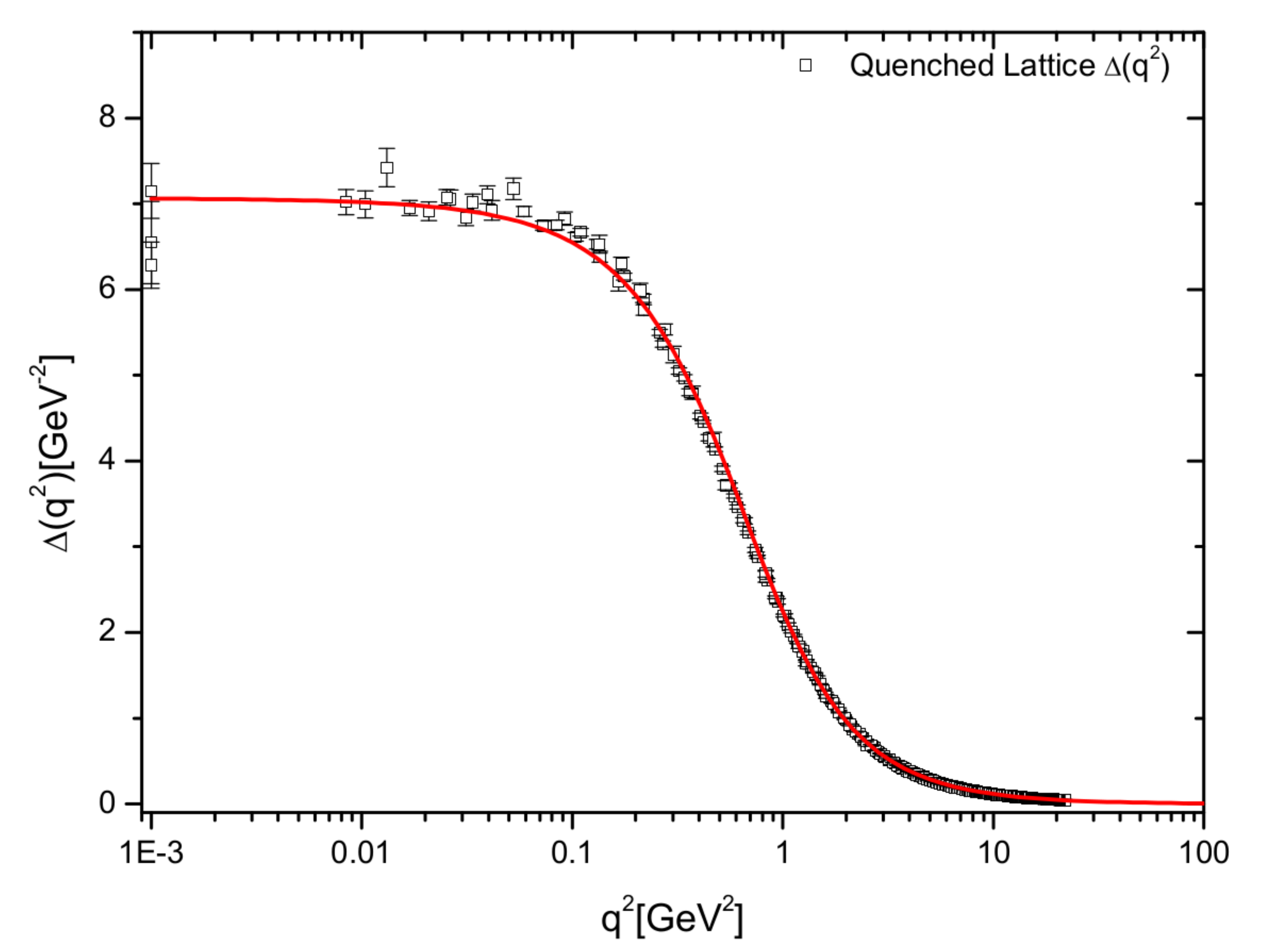}
\end{center}
\vspace*{-0.20cm}
\caption{\label{fig:QCD-DSEs}
Figures adapted from Ref.~\citenum{Bhagwat:2003vw} and~\citenum{Papavassiliou:2015aga}, respectively.
{\emph Left-panel:} Data points are lattice-QCD results for $M(p^2)$ from Ref.~\citenum{Bowman:2002kn}; solid curves are solutions of the quark DSE (gap equation) for $M(p^2)$, obtained using the current-quark mass that best fit lattice-QCD results; and the (red) lowest solid curve represents the gap equation's solution in the chiral limit.
{\emph Right-panel:} The quenched lattice gluon propagator $\Delta(q^2)$ and an adapted solution }
\end{figure}

\section{Dyson-Schwinger equations}

The Green functions of a quantum field theory contain all its physical properties and one can state that the theory is solved once every Green function is determined. During the past years, a cross-fertilization between QCD-DSEs and lattice-QCD has provided further insight into the non-perturbative structure of the lowest $n$-point Green functions: quark, gluon and ghost propagators as well as quark-gluon, three-gluon and four-gluon vertices.

This section is focused on highlighting the results recently delivered for the quark and gluon propagators, in Landau gauge. Note, however, that studies of the infrared point-wise behavior for the ghost propagator~\cite{Aguilar:2008xm}, the quark-gluon vertex~\cite{Aguilar:2016lbe}, and the three- and four-gluon~\cite{Eichmann:2014xya, Eichmann:2015nra} vertices have recently released too. Another remarkable achievement of this kind of studies has been the combination of the ghost and gluon propagators in order to deliver a process-independent running-coupling for QCD which saturates at zero momentum~\cite{Binosi:2016nme, Zafeiropoulos:2019flq}.

The dressed-quark propagator in Landau gauge can be written in the following form:
\begin{equation}
S_f(p) =  -i \gamma\cdot p\, \sigma_V^f(p^2) + \sigma_S^f(p^2) = 1/[i\gamma\cdot p\, A_f(p^2) + B_f(p^2)]\,.
\label{eq:Sq}
\end{equation}
It is known that for light-quarks the wave function renormalisations, $Z_f(p^2)=1/A_f(p^2)$, and dressed-quark masses, $M_f(p^2)=B_f(p^2)/A_f(p^2)$, receive strong momentum-dependent corrections at infrared momenta~\cite{Bhagwat:2003vw}: $Z_f(p^2)$ is suppressed and $M_f(p^2)$ enhanced (see left-panel of Fig.~\ref{fig:QCD-DSEs}). These features are an expression of dynamical chiral symmetry breaking (DCSB) and, plausibly, of confinement via the violation of reflection positivity.

The gluon propagator in Landau gauge assumes the totally transverse form
\begin{equation}
i \Delta_{\mu\nu}(q) = -i P_{\mu\nu}(q) \Delta(q^2); \quad\quad P_{\mu\nu}(q) = g_{\mu\nu} - q_\mu q_\nu / q^2 \,,
\end{equation}
where the scalar form factor $\Delta(q^2)$ is related to the all-order gluon self-energy. Lattice-QCD simulations~\cite{Bogolubsky:2009dc, Bowman:2007du, Cucchieri:2007rg, Cucchieri:2009zt} have recently confirmed~\cite{Cornwall:1981zr} that this form factor saturates in the deep infrared (see right-panel of Fig.~\ref{fig:QCD-DSEs}) indicating gluon mass generation. A demonstration of how this occurs at the level of the gluon DSE is given in Ref.~\citenum{Aguilar:2008xm}.

The transition from a massless to a massive gluon propagator can be implemented as $\Delta^{-1}(q^2) = q^2 J(q^2) + m^2(q^2)$, where $J(q^2)$ is the gluon's dressing function and $m^2(q^2)$ is the momentum dependent gluon mass, dynamically generated by the Schwinger mechanism~\cite{Schwinger:1962tn, Schwinger:1962tp}. This can be summarize as follows: the vacuum polarization of a gauge boson that is massless at the level of the original Lagrangian may develop a massless pole, whose residue can be identified with $m^2(0)$. The origin of the aforementioned poles is due to purely non-perturbative dynamics. They act as composite Nambu-Goldstone bosons which are colored, massless and have a longitudinal coupling. These features maintain gauge invariance and makes them disappear from any on-shell $S$-matrix element.


\begin{figure}[!t]
\begin{center}
\includegraphics[width=0.80\textwidth, height=0.15\textheight] {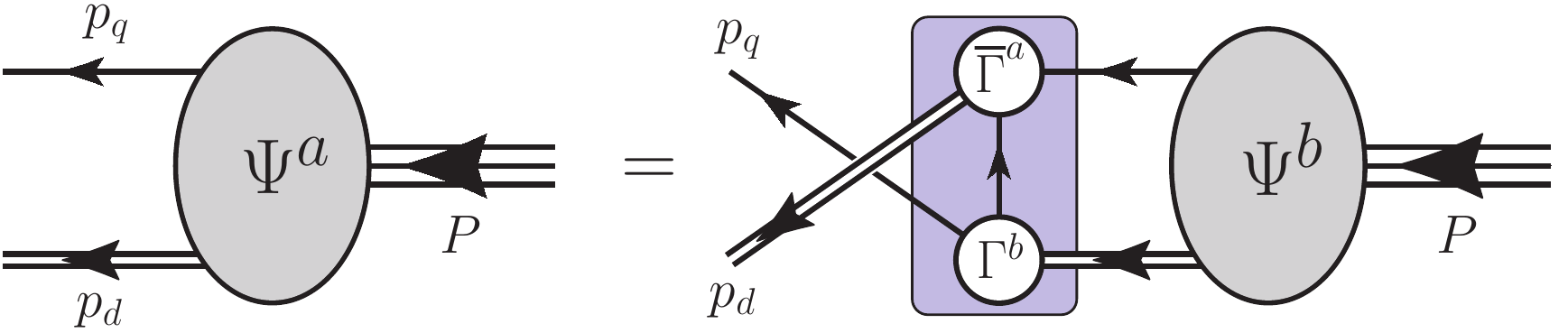}
\end{center}
\vspace*{-0.20cm}
\caption{\label{fig:Faddeev} Poincar\'e-covariant Faddeev equation: a homogeneous linear integral equation for the matrix-valued function $\Psi$, being the Faddeev amplitude for a baryon of total momentum $P = p_q + p_d$, which expresses the relative momentum correlation between the dressed-quarks and -diquarks within the baryon. The (purple) highlighted rectangle demarcates the kernel of the Faddeev equation: {\it single line}, dressed-quark propagator; {\it double line}, diquark propagator; and $\Gamma$, diquark correlation amplitude.
}
\end{figure}

\section{Covariant bound-state equations}

Mesons appear as free-poles in the quark-antiquark scattering matrix and baryons in the complete three-quark one. Their properties are extracted upon solving homogeneous bound-state equations (Bethe-Salpeter for mesons and Faddeev for baryons) which are valid at pole positions and need information of the lowest $n$-point Green functions of QCD. In combination they provide a powerful tool to calculate experimentally accessible hadron observables, {\emph e.g.} meson and baryon masses, decay constants, scattering processes, and electromagnetic properties such as form factors.

The problem of solving the Faddeev equation can be transformed into that of solving the linear, homogeneous matrix equation depicted in Fig.~\ref{fig:Faddeev}. This is because two decades of studying baryons as three-quark bound-states~\cite{Oettel:1998bk, Eichmann:2016yit, Lu:2017cln, Yin:2019bxe} have demonstrated the appearance of soft (non-pointlike) fully-interacting diquark correlations within baryons, whose characteristics are greatly influenced by DCSB~\cite{Segovia:2015ufa}. Note that a baryon described by Fig.~\ref{fig:Faddeev} can be interpreted as a Borromean bound-state where the binding energy is given by two main contributions: One part is expressed in the formation of tight diquark correlations, the second one is generated by the quark exchange depicted in the highlighted rectangle of the Fig.~\ref{fig:Faddeev}. This exchange ensures that no quark holds a special place because each one participates in all diquarks to the fullest extent allowed by its quantum numbers. The continual rearrangement of the quarks guarantees that the wave function complies with the baryon's fermionic nature.

\begin{figure}[!t]
\begin{center}
\includegraphics[width=0.45\textwidth, height=0.25\textheight] {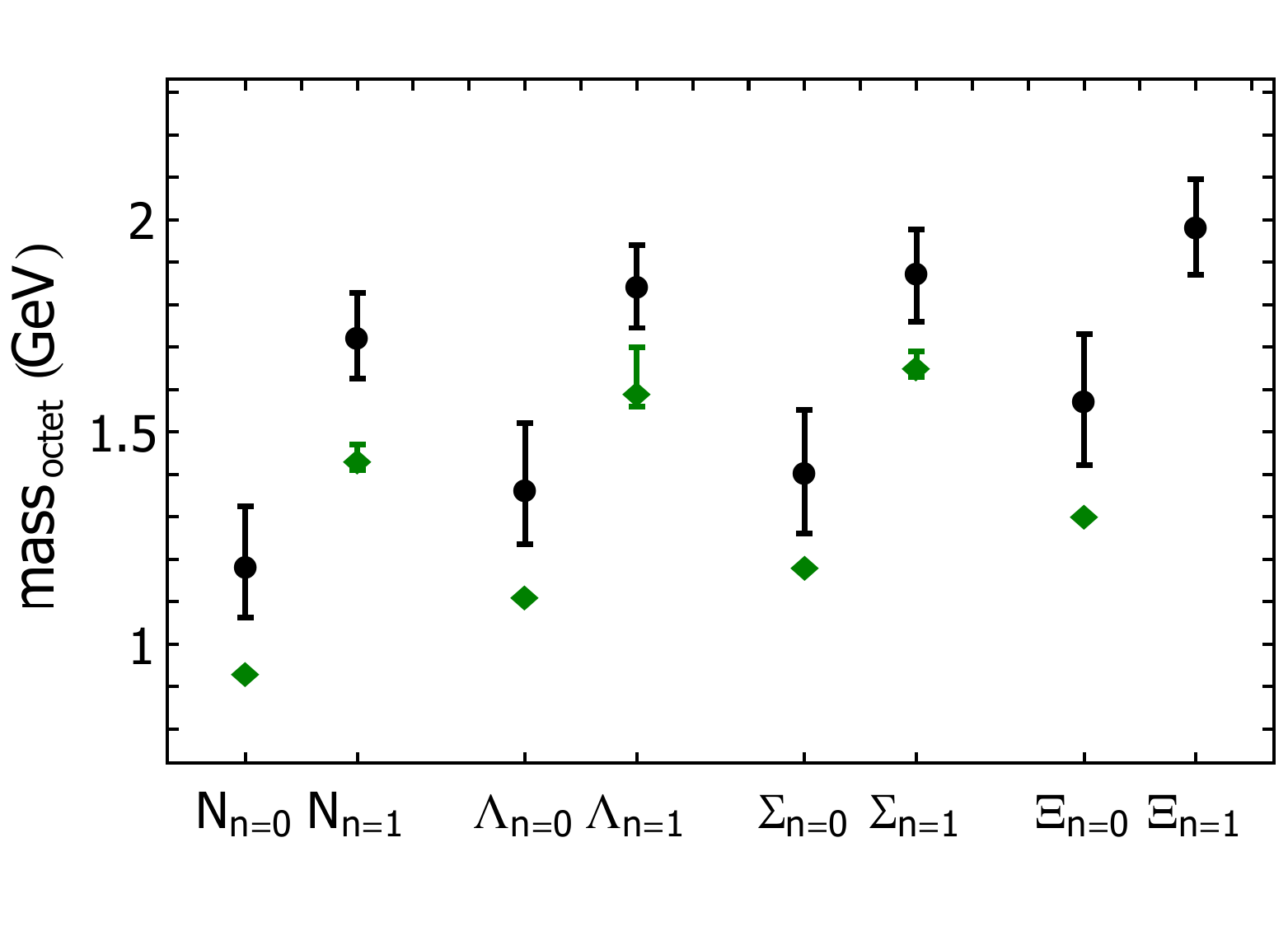}
\hspace*{0.25cm}
\includegraphics[width=0.45\textwidth, height=0.25\textheight] {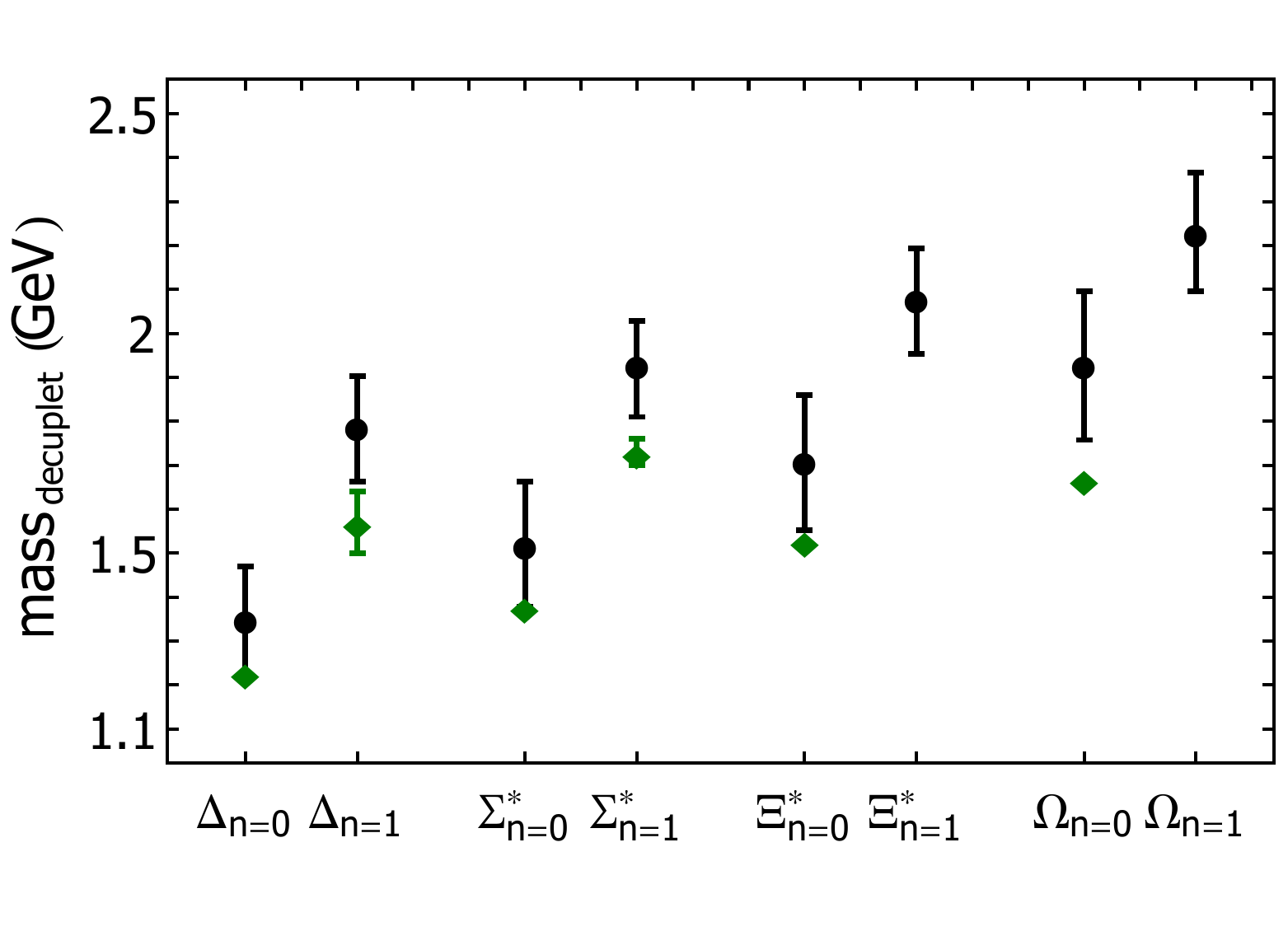}
\end{center}
\vspace*{-0.50cm}
\caption{\label{fig:Spectrum} (Black) solid points are the computed masses for the ground- and first-excited state of the octet (left-panel) and decuplet (right-panel) baryons~\cite{Chen:2019fzn}. The vertical riser indicates the response of our predictions to a coherent $\pm 5$\% change in the mass-scales that appear in the baryon bound-state equation. The horizontal axis lists a particle name with a subscript that indicates whether it is ground-state ($n=0$) or first positive-parity excitation ($n=1$). (Green) diamonds are empirical Breit-Wigner masses take from Ref.~\citenum{Tanabashi:2018oca}, the estimated uncertainty in the location of a resonance's Breit-Wigner mass is indicated by an error bar.}
\end{figure}

Figure~\ref{fig:Spectrum} shows the computed masses of octet and decuplet baryons and their first positive-parity excitations~\cite{Chen:2019fzn}. It is apparent that the theoretical values are uniformly larger than the corresponding empirical ones. This is because our results should be viewed as those of a given baryon's dressed-quark core, whereas the empirical values include all contributions, including meson-baryon final-state interactions (MB\,FSIs), which typically generate a measurable reduction~\cite{Suzuki:2009nj}. This was explained and illustrated in a study of the nucleon, its parity-partner and their radial excitations~\cite{Chen:2017pse}; and has also been demonstrated using a symmetry-preserving treatment of a vector$\,\times\,$vector contact interaction \cite{Lu:2017cln, Yin:2019bxe}. Identifying the difference between our predictions and experiment as the result of MB\,FSIs, then one finds that such effects are fairly homogeneous across the spectrum. Namely, they act to reduce the mass of ground-state octet and decuplet baryons and their first positive-parity excitations by roughly $0.25\,$GeV.

It is important to highlight here that a possible way to evade the effects of MB\,FSIs is studying electromagnetic transition form factors of nucleon resonances because their contribution to this kind of observables is soft and disappears quickly, allowing to explore the dressed-quark core of a baryon when electromagnetic proves of high-momenta are used~\cite{Segovia:2014aza, Segovia:2015hra, Segovia:2016zyc, Chen:2018nsg, Lu:2019bjs}.


Herein, I have sketched the exploitation of QCD-DSEs to baryon physics. There are also numerous applications to topical problems in the meson sector. For example, programmes are approved at JLab12, proposed at the CERN-SPS, and possible at the electron-ion collider which would reveal much about parton distributions within QCD's Nambu-Goldstone modes~\cite{Aguilar:2019teb}. A strong motivation for such measurements is the fact that the leading $x$-moment of the pion's (kaon's) GPD provides access to the distributions of mass and momentum within the pion (kaon). Since these distributions can be calculated~\cite{Frederico:2009fk, Mezrag:2014jka}, such data can significantly influence future theoretical perspectives.

\begin{figure}[!t]
\begin{center}
\includegraphics[width=0.65\textwidth] {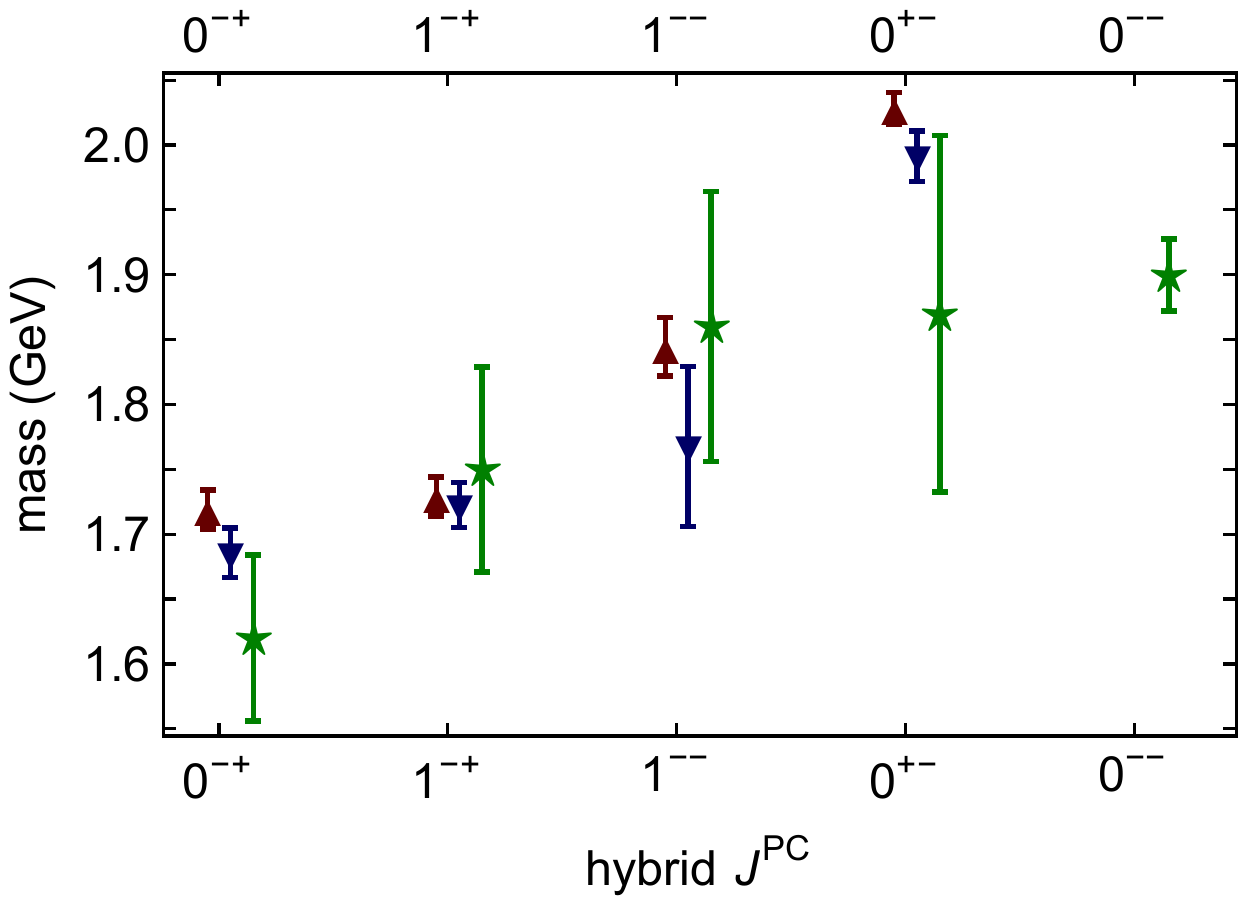}
\end{center}
\vspace*{-0.20cm}
\caption{\label{fig:Spectrum2} Figure taken from Ref.~\citenum{Xu:2019sns}. Light hybrid-meson spectrum computed with QCD-DSEs (stars, green), and the re-scaled lattice-QCD results.}
\end{figure}

Finally, QCD-DSEs can also be applied to exotic hadrons; namely, tetra- and penta-quark bound-states~\cite{Fischer:2017cte} but also systems with valence glue, {\emph e.g.} hybrid mesons~\cite{Xu:2019sns}, $Q\bar{Q}G$ (see Fig.~\ref{fig:Spectrum2}); hybrid baryons, $QQQG$; and even glueballs~\cite{Souza:2019ylx}, $GG$. Here, $G$ is a nebulously defined ``constituent gluon" degree of freedom, whose nature will only be known once such systems are detected.

\section{Perspective}

The Dyson-Schwinger equations provide a fully relativistic continuum approach for the calculation of hadron properties. Among its advantages, one can highlight that it provides access to infrared and ultraviolet momenta, and covers the full quark mass range from the chiral limit to arbitrarily large current masses. At the present stage, especially in the baryon sector, we are only beginning to explore its potential and possibilities.


\section*{Acknowledgments}

%
Work supported by: Ministerio de Economia Industria y Competitividad (MINECO), under grant no. FPA2017-86380-P; Helmholtz International Center for FAIR within the LOEWE program of the State of Hesse and by the DFG grant FI 970/11-1; National Natural Science Foundation of China, grant nos. 11535005, 11805097; Jiangsu Province Natural Science Foundation grant no. BK20180323; and Jiangsu Province Hundred Talents Plan for Professionals.



\bibliographystyle{ws-procs9x6}
\bibliography{Procs-Hadron2019_JSegovia}

\end{document}